1

# New Software Metrics for Evaluation and Comparison of Advanced Power Management Systems

Dr. J. Chris Foreman Ph.D., *Member IEEE*, Dr. Rammohan K. Ragade Ph.D., *Senior Member IEEE*, and Dr. James H. Graham Ph.D., *Senior Member IEEE*

**Abstract** – A set of software metrics for the evaluation of power management systems (PMSs) is presented. Such systems for managing power need to be autonomous, scalable, low in complexity, and comprised of portable algorithms in order to be well applied across the varying implementations that utilize power systems. Although similar metrics exist for software in general, their definitions do not readily lend themselves to the unique characteristics of power management systems or systems of similar architecture.

**Index Terms** – Software, Power, Management, Metrics, Portability, Scalability, Complexity, Autonomy, Visualization

## I. INTRODUCTION

Power Management Systems (PMSs) are utilized in an ever-increasing variety of devices and missions. These systems are typically implemented in software and are responsible for keeping the power system operating reliably and efficiently. The varying missions of PMSs scale from portable devices and hybrid / electric vehicles, to power generating plants and large industrial processes. While there is much research in all these areas, algorithms developed for one mission are often not designed to be portable and scalable to another mission that is very different in size. Complexity is also an issue since small devices do not always have the available processing power and memory to accommodate a large power management footprint. Likewise, large industrial and power generating processes are often controlled by legacy hardware that may not provide sufficient functionality for complex algorithms. Finally, autonomy is a necessary characteristic as it is often impractical to have the end user perform power management functions, or when minimizing these user interactions is desirable.

The software metrics presented evaluate intelligent PMS software across these characteristics of scalability, portability, complexity, and autonomy in order to compare algorithms and encourage cooperative development efforts across a wider span of implementations.

### A. Previous work
Autonomy is a measure of the system's ability to make decisions and perform the mission at hand with minimal human intervention. Clough [1] illustrates the problems of measuring autonomy in unmanned aerial vehicles (UAVs) using existing definitions and comes up with a new procedure. Sholes [2] further evolves the work of Clough, again on UAV control systems, with application to various implementations. Rushby and Crow [3] evaluated automated fault detection and recovery in expert systems for a manned maneuvering unit. Portability and scalability in PMSs are particularly difficult, yet important, because a broader range of application environments exist, ranging from ladder logic in PLCs to embedded-C on custom processors. Complexity is a well documented software metric as in decision points, e.g. McCabe's cyclomatic complexity [5] and Henry and Selig's structural complexity [6], and input and output counts, e.g. Henry and Kafura's information flow [7]. However, it is important to gauge readability when considering management systems, e.g. as in rule-based expert systems by Chen and Suen [4]. Good readability will ensure the PMS is easily understood and maintained by personnel, which is critical in many control environments.

### B. Proposed solution
New metrics for measuring portability, scalability, complexity, and autonomy are presented to characterize advanced power managements systems. These metrics complement the typical mission-level metrics utilized for a common PMS, e.g. power efficiency, power handling capability, etc. As power becomes an increasingly critical resource, good management is needed for optimal utilization and preservation. This leads to the development of software algorithms and, subsequently, the need for software metrics to evaluate these algorithms that are meaningful to the mission of power management. The use of these metrics encourages the development of PMSs whose software components are both more effectively utilized and more capable of being reused. The metrics are designed to be easily computable by non-software engineers and radar graphs are proposed to allow quick visualization of the resulting metrics. The visualization capabilities also enable software changes to be tracked by the metrics and compared among multiple PMS implementations.

## II. DEFINITION OF METRICS

When evaluating the power management software, the code should be modularized, i.e. broken into fundamental modules, for a granular analysis. As appropriate for a given architecture, algorithms can be handled individually, or as modules such as rule sets, equation sets, and/or neural networks.

The metrics are defined here and will be applied and interpreted in Section III.

*A. Portability and scalability*

Portability $P$ is a measure of the development effort required to move software among similar PMSs. This is a lateral or horizontal move but with similar functional scope. This is in contrast to scalability $S$, which seeks to measure the development effort required to add new scope and functionality to an individual PMS. Table 1 defines how individual modules can be assessed while (1) defines how portability would be summed for all modules in a PMS – likewise for scalability substituting $S$ and $s$ for $P$ and $p$ in (1). If appropriate, the evaluator can choose a value between two levels, e.g. 2.5, to improve the granularity of the classification. A weighting factor $w_i$ is included to increase the significance of individual modules if appropriate.

Table 1. Portability and scalability factors.

| Portability factor $p_i$ | Scalability factor $s_i$ | Degree of software changes required |
|---|---|---|
| 3 | 3 | No change |
| 2 | 2 | Parameter-level changes |
| 1 | 1 | Code-level changes |
| 0 | 0 | Not portable / scalable |

$$P = \frac{\sum_{i}^{N} w_i p_i}{\sum_{i}^{N} w_i} \quad \text{also with } S,s \text{ for } P,p \qquad (1)$$

Although portability and scalability differ as defined above, together they represent the modularity and modifiability of the PMS. These quantities can be drawn together on a linear radar graph in opposing direction, where total length is a measure of the modularity and modifiability and the center location indicates the dominant quantity between portability and scalability. An example of this is illustrated in Fig. 1 where the center is shifted towards scalability.

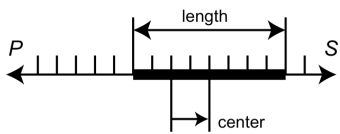

Figure 1. Portability and scalability graph.

*B. Complexity*

As previously noted, complexity is a well-defined metric in software science but needs to include readability and inputs and outputs (I/O) when applied to PMSs in control system environments, and perhaps control systems software in general. Complexity for an individual software module is defined by (2). Readability $r$ is defined in Table 2 according to how easily a human reader can understand the code. As in portability and scalability, values between two levels can be selected if appropriate to improve the granularity, e.g. modules comprised of a combination of equations and procedures for a readability of 2.5. Readability is important because PMSs need to be periodically maintained by several users, sometimes spanning years between edits. Also, PMSs that automate an operator's actions may not get good buy-in or acceptance if they and their actions are not well understood. The next term, McCabe's complexity [5] represented by $m$, is the number of decision points plus 1. Fan-in $f_{in}$ and fan-out $f_{out}$ are included to account for the field inputs typical in control systems. Brooks [8] and Belady [9] have proposed squaring of the product of fan-in and fan-out, however this can quickly result in a very high number that obscures the importance of the other factors of McCabe and readability. Therefore, this term is left un-squared due to the nature of PMSs of typically handling high I/O counts.

Table 2. Readability factor in complexity.

| Readability factor $r_i$ | Meaning |
|---|---|
| 1 | Natural language or simple rule statements allowing straightforward interpretation of meaning. |
| 2 | Computable equation that requires computing equations to determine meaning. |
| 3 | Procedural computation that requires following a difficult procedure to determine meaning. |

$$c_i = r_i m_i (f_{in} f_{out}) \qquad (2)$$

The modular complexities defined by $c_i$ in (2) are summed for the whole PMS as with portability and scalability by (1). This complexity metric can be added to the previous radar graph in Fig. 1 for portability and scalability to form a surface plot for evaluation. Fig. 2 illustrates how these three metrics are quickly observed with portability and scalability values stretching the quality of the PMS upwards and complexity pulling the surface downwards.

The values for $c_i$ and subsequently for the PMS's complexity can vary widely for dissimilar applications since the specific algorithms and I/O counts affect these dramatically. Therefore, complexity is best utilized when comparing architectures within a given mission in consideration of selecting candidate PMSs from a range of possibilities.

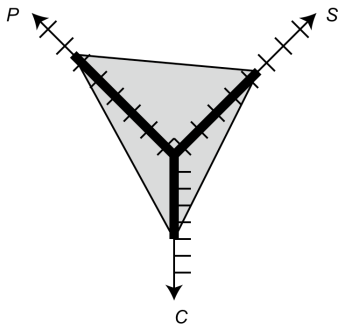

Figure 2. Complexity graph with portability and scalability.

*C. Autonomy*

Achieving autonomy frees the operator from control tasks, handles trouble conditions automatically, allows strategic decisions to be automated, and finally enables cooperation with peer systems within the environment for a coordinated solution. Unlike the previous metrics, autonomy is measured on the whole PMS as opposed to fundamental modules. In PMSs, the key parameters are:

- Operator independence – requiring minimal user interaction, having automation.
- Self-preservation – the ability to handle trouble conditions automatically, recover and continue the mission, and fail in a safe manner.
- Strategy – the ability to enhance the control of the power system and thus add to its capabilities.
- Coordination – the ability to cooperate with other users and PMSs.

Operator independence $A_I$ is determined by measuring the percentage of manual tasks previously performed by the operator that will be now performed by the PMS as in (3).

$$A_k = \frac{\sum tasks_{auto}}{\sum tasks_{total}} \cdot 100\% \quad \text{where } k = I, P, S, C \quad (3)$$

Likewise with self-preservation $A_P$, the number of trouble conditions or alarms now automatically handled by the PMS is measured as a percentage of the total system alarms or total number of failure modes. This is calculated similarly to $A_I$ as in (3).

Strategic capability is a qualitative measure of how well the PMS enhances the whole system's ability to handle power efficiently and effectively. Table 3 outlines a guide for assigning a value for $A_S$ where 0% is no effort to enhance and 100% is the theoretical limit for the given PMS. Within each meaning, a range is given to allow the evaluator to score the performance of the specific system.

Table 3. Strategy.

| Strategy $A_S$ | Meaning |
|---|---|
| 75% - 100% | Many new goals or strategies applied to enhance the PMS capabilities to the theoretical limit |
| 50% - 75% | Multiple new goals or strategies applied to enhance the PMS capabilities, multi-goal optimization |
| 0% - 25% | One new goal or strategy applied to enhance the PMS capabilities, single-goal optimization |
| 0% | No enhancement |

Similarly to $A_S$ coordination is also a qualitative quantity that measures how well the PMS interacts with its environment. As before, a range is given to allow the evaluator to score the performance of the specific system. Table 4 defines how the value for $A_C$ is assigned.

Table 4. Coordination.

| Coordination $A_C$ | Meaning |
|---|---|
| 75% - 100% | Full cooperation with all entities, intuitive to theoretical limit. |
| 50% - 75% | Limited coordination with other systems and/or coordination of the influence of multiple users. |
| 0% - 25% | Aware of other systems but little or no coordination. May recognize multiple users. |
| 0% | Unaware of other systems. Only operator-level control by one user at a time. |

Once the above four sub-metrics for autonomy are determined, they can be plotted on a radar graph for analysis and comparison. This allows a quick graphical interpretation of autonomy in the PMS where height and width correspond to automation and intelligence and the total area is the total autonomy metric. Indeed, a good definition of autonomy could be described generally by (4) where the multidimensional quantities of automation and intelligence are combined to characterize autonomy. The metrics defined in this paper relevant to these are demonstrated in Fig. 3.

$$\overrightarrow{Autonomy} = \overrightarrow{Automation} \times \overrightarrow{Intelligence} \quad (4)$$

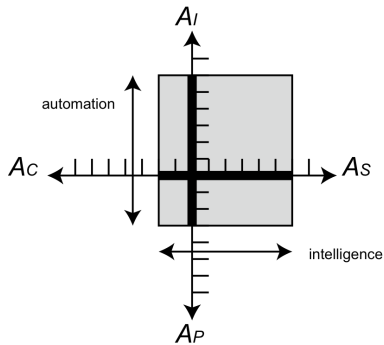

Figure 3. Autonomy.

III. APPLICATION OF METRICS

Determination of the portability, scalability, and complexity metrics are done on a module-by-module basis and summed according to the equations in Section II. For PMSs that are likely to be implemented in process control systems, the typical program structure includes rules, equations, and simple glue code, as well as model-based elements such as neural networks.

*A. Portability and scalability metrics example*

Portability and scalability will likely be a qualitative decision by a software developer according to the guidelines in Table 1. Therefore, it is advantageous to break the PMS into the simplest modules practical. Modules can be assessed fractional values between those of Table 1 if deemed appropriate, e.g. level 2.5 for very minor parameter changes and 0.5 for very significant codes changes. Table 5 demonstrates the assessment of a set of modules.

Table 5. Portability and scalability metric application.

| Module (assumed equal weight) | Portability to another instance of a similar PMS | Scalability to add new scope to an existing PMS |
|---|---|---|
| 1 | Changed very few parameters, level 2.5 | No changes, level 3 |
| 2 | Changed parameters and some minor code changes, level 1.5 | Changed parameters and several lines of code, level 1 |
| 3 | No changes, level 3 | New module needed to be added to handle new scope, level 0 |
| 4 | Changed parameters, level 2 | Changed very few parameters, level 2.5 |
| Total by (1) | 2.25 | 1.625 |

*B. Complexity metric example*

Rules are typically in the form of IF-THEN or SWITCH-CASE type formats, especially in legacy industrial systems and limited capability micro-systems. Rules therefore take the form of:

```
IF A<K1 THEN SET X=10 ELSE X=B;
F(A) = A * SIN (K3*T) + K2
IF F(A)>K4 THEN SET C=TRUE ELSE SET C=FALSE;

SWITCH (X) {
    CASE X<=0: SET MOTOR=OFF;
    CASE X<=10: SET MOTOR=LOW;
    CASE X>10: SET MOTOR=HI;
    DEFAULT: SET MOTOR=ERROR;
}
```

If the code block above represented a module to be evaluated for complexity, the results would be: 2 IF-THEN statements and 3 switch-case decisions for 5 decisions total; and 1 equation.

$r = 1.25$; 75% level 1 and 25% level 2 readability
$m = 6$; #decision points + 1
$f_{in} = 2$; inputs A,B (K's are constants)
$f_{out} = 2$; outputs C,MOTOR (X internal variable)
$c = 30$; per equation (2)

Neural networks and other black-box elements are measured based only on their I/O count and difficulty with internal readability. For a neural network, readability is $r=3$ by Table 2 since the only way to determine the output a neural network is to provide an input. There are no decision points, i.e. the input pattern always flows through to subsequent the output pattern, therefore $m=1$. For a small neural network of 10 inputs and 4 outputs, the resulting complexity becomes $c=3*1*10*4=120$.

*C. Autonomy metric example*

The autonomy metric is applied to the whole PMS based on the definitions outlined in Section II. An example application to an existing PMS is the software-agent-based approach by Foreman and Ragade [10] at a hydro-generating plant. The plant is a run-of-the-river, Kaplan turbine design with three units rated 25MW each at full load. The software agents negotiated water flow among the units and automated several operator tasks.

Table 6 demonstrates this metric application and Fig. 4 illustrates the autonomy metric radar graph and area calculation by (4).

Table 6. Autonomy metric for the hydro PMS.

| | Evaluation | Value |
|---|---|---|
| $A_I$ | By handling the two main operator inputs to control a unit, being Kaplan blade tilt and wicket gate position, the units are theoretically fully automated by the PMS. | 100% |
| $A_P$ | Outside of extraordinary trouble conditions, the typical conditions handled by the operator were cavitation, vibration, and generator temperature excursions. These are fully handled by the PMS. | 100% |
| $A_S$ | The PMS implements a few strategies for operating optimization but there is room for additional strategies. Assessed by Table 3. | 75% |
| $A_C$ | The PMS is designed to handle multiple users and be aware of the other units. There is some room for more intuitive interaction with other units and even units at other plants, e.g. river-level coordination. Assessed by Table 4. | 75% |
| Total by (4) | | 3.0 |

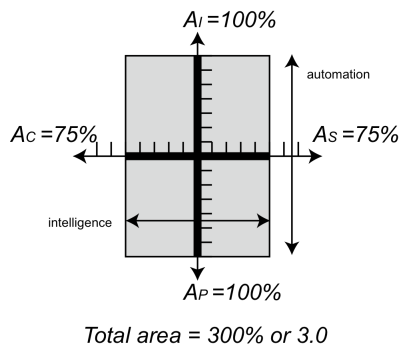

Figure 4. Autonomy in the hydro PMS.

*D. Visualization of PMS changes and comparisons*

Once a graphical representation of these metrics is developed, both proposed changes to an existing PMS and comparison with a peer PMS can be performed visually. This enables a clear interpretation of differences across a wider audience and facilitates good participation in system planning. Fig. 5 illustrates how a modification to an existing PMS that changes autonomy can be visualized. In this figure, the self-preservation metric has been improved from the first PMS to the second PMS, viewing from left to right.

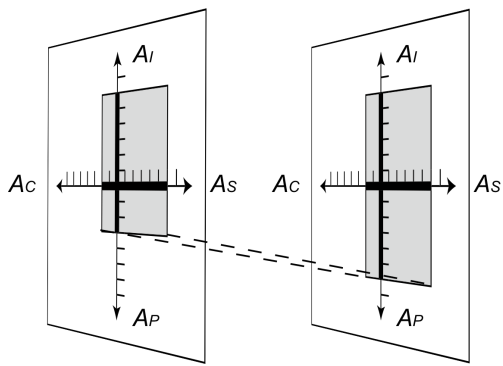

Figure 5. Autonomy changes during PMS modification.

Fig. 6 illustrates how multiple PMSs can be visually compared together. In this figure, autonomy is again compared across three candidate PMSs to determine which has the desired profile for the current mission, viewing from left to right. The second PMS improves cooperation at the expense of operator independence compared to the first PMS. The third PMS has somewhat less improved cooperation and improved self-preservation without a loss in the other metrics compared with the first PMS.

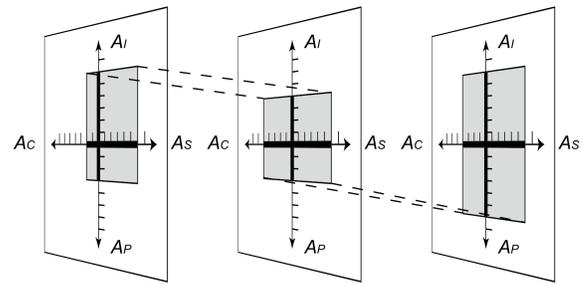

Figure 6. Autonomy metrics for three competing PMSs.

## IV. CONCLUSIONS

The metrics defined for power management systems provide a framework for evaluating the performance and maintainability of such systems. They are designed such that individuals can apply and interpret them with little software science experience. This encourages PMS designers to include these metrics in their PMS proposals and research due to the relatively small additional work required for their incorporation. In particular, the radar graph visualization allows quick understanding by lay people for presentation purposes. These metrics are valuable in their application to power management systems due to the lack of specialized metrics in this area and the difficulty in applying incompatible metrics, e.g. the autonomy metrics mentioned in Section I for UAVs that focus on mobility.

BIOs

**J. Chris Foreman** (Ph.D. Computer Science and Engineering degree, University of Louisville, 2008) is a member of IEEE, the Power and Energy Society, and also holds both B.S. (1990) and M.Eng. (1996) degrees in Electrical Engineering from the University of Louisville. He is a postdoctoral associate at the University of Louisville in Louisville, KY and performs research in SCADA security, renewable energy systems, and smart power grids. He has worked primarily in the power generation industry, among others, in industrial process control since 1993. Specializing in advanced control techniques and processes, he has managed several projects to improve production, efficiency, and reduce emissions. He has worked for companies such as Westinghouse Process Control Division (now Emerson Process Management), Cinergy (now Duke Energy), and Alcoa Inc.

**Rammohan K. Ragade** (Ph.D., I. I. T. Kanpur, India (1968)) is a Professor of Computer Engineering and Computer Science at the University of Louisville. He holds a B.E. degree in Electrical Power Engineering from I. I. Sc. Bangalore, India (1964). He served as the Coordinator for the Ph.D. Program in Computer Science and Engineering from 1999-2005. He has written well over 100 papers, including journal articles, refereed conference papers, chapter contributions to books and is the co-editor of four books. He is a senior member of IEEE. He is a member of the ACM. He has taught graduate courses in Software Engineering and Advanced Software Engineering, Software Design, Computer Security, Knowledge Engineering, Computer Architecture, and Simulation Modeling. His research interests include agent technologies, object oriented methodologies, real-time modeling, human computer interaction, knowledge engineering and rule-based expert systems, and system simulation. He has held and participated in several funded research grants and contracts.

**James H. Graham** (Ph.D. degree, Purdue University, 1980) is the Henry Vogt Professor and the Chair of Electrical and Computer Engineering at the University of Louisville in Louisville, KY. He also received his Bachelor's degree in Electrical Engineering from the Rose-Hulman Institute of Technology and the M.S. degree from Purdue University in 1978. He is a senior member of the Institute of Electrical and Electronics Engineers (IEEE) and a registered professional engineer. He has over thirty years of experience in the computer engineering and electrical engineering fields. Prof. Graham has served as a faculty member at Rensselaer Polytechnic Institute and as a product engineer with General Motors Corporation. His research interests involve information security, algorithms for computational science, intelligent systems, distributed computing, computer simulation, and intelligent energy systems.